\documentclass[lettersize,journal]{IEEEtran}
\usepackage{amsmath,amsfonts}
\usepackage{siunitx}
\usepackage{algorithmic}
\usepackage{algorithm}
\usepackage{array}
\usepackage[caption=false,font=normalsize,labelfont=sf,textfont=sf]{subfig}
\usepackage{textcomp}
\usepackage{stfloats}
\usepackage{url}
\usepackage{verbatim}
\usepackage{graphicx}
\usepackage{cite}
\usepackage{graphicx}
\usepackage{tikz}
\usepackage[hidelinks]{hyperref}
\begin{document}

\bstctlcite{IEEEexample:BSTcontrol}

\title{Investigation of Shock Wave Dynamics in Complex Plasma via Computational Modelling}

\author{Prateek Lamoria, Anton Kananovich and Surabhi Jaiswal,~\IEEEmembership{Member,~IEEE}}

\maketitle

\begin{abstract}
Piston-driven shock waves in dusty plasma monolayers have been observed experimentally and studied using molecular dynamics (MD) simulations. However, previous MD simulations were restricted to strictly two-dimensional geometries with periodic or reflecting boundaries and therefore could not capture the out-of-plane buckling of microparticles observed experimentally under shock compression. We present three-dimensional MD simulations of a piston-driven shock in a two-dimensional dusty plasma. The model incorporates finite-harmonic vertical confinement, fixed boundaries, Epstein drag and a microparticle size distribution matched to previous experiments. The simulations recover the experimentally observed linear scaling between the shock and piston Mach numbers and, to the best of our knowledge, reproduce, for the first time in MD, shock-induced buckling of the monolayer. These results bridge a long-standing gap between laboratory observations and simulations of two-dimensional dusty plasma shocks, and provide a validated framework for investigating out-of-plane and wake-mediated shock physics in strongly coupled systems.
\end{abstract}


\section{Introduction}

\IEEEPARstart{A}{shock} wave is a thin, dissipative, non-linear wave across which hydrodynamic quantities such as pressure and density change rapidly. It propagates supersonically (with $v>c_s$), abruptly compressing, heating, and decelerating the medium that flows into it~\cite{choudhuri1998fluidsplasmas}. Shock waves occur in a variety of media, such as neutral fluids~\cite{zeldovich2002shockwaves}, condensed matter~\cite{forbes2012shockcompression}, and plasmas~\cite{gurnett2005plasma}. They have also been widely observed in space plasmas, including protostellar flows associated with the early stellar evolution~\cite{annurev:/content/journals/10.1146/annurev.astro.39.1.403,refId0} and in the interaction of the supersonic solar wind with Earth’s magnetosphere~\cite{https://doi.org/10.1029/JZ067i010p03791,Burgess2005-za}.

Dusty or complex plasmas (plasmas containing charged microparticles) are ubiquitous throughout the universe. In the laboratory, dusty plasmas offer a unique opportunity to study a variety of collective plasma phenomena with exquisite detail at the microscopic level~\cite{feng2007accurate,feng2011errors,mohr2019algorithms,PhysRevE.111.045201,kananovich2026large, PhysRevE.93.041201, 10.1063/1.5040417, nxt3-qwj6, PhysRevResearch.6.013119}. Because microparticles acquire large electric charges, large-amplitude compressional waves can be excited readily in dusty plasmas. Under suitable conditions, these waves become nonlinear and can steepen into shock waves~\cite{https://doi.org/10.1002/ctpp.200910022,PhysRevE.79.055401}. For shock waves generated in the laboratory dusty plasmas, the relevant longitudinal sound speed ($c_l$) is typically a few centimetres per second~\cite{10.1063/1.4960032,PhysRevE.101.043211}, while the shock front widths are of the order of a few millimetres for laboratory dusty plasma~\cite{PhysRevE.104.055201}. Thus, these strongly coupled systems are quite convenient for investigating shock wave features.

There is a rich literature available on the study of linear and non-linear modes and coherent structures in a complex plasma, where dust ion-acoustic shock waves (DIASWs), dust-acoustic shock waves (DASWs), and dust-lattice shock waves (DLWs) have been extensively studied both theoretically and experimentally over the past couple of decades~\cite{10.1063/1.4960032, PhysRevE.100.043203, PhysRevLett.118.025001, PhysRevE.101.043211, PhysRevLett.108.065004, PhysRevE.104.055201}.

The interaction between charged dust particles in many low-temperature plasma experiments can be approximated by the screened Coulomb potential (also known as the Debye-Hückel and Yukawa potential)~\cite{konopka2000measurement}, given as

\begin{equation}
\Phi(r_{ij}) = \left(\frac{Q_iQ_j}{4\pi \epsilon_0 r_{ij}}\right)\exp \left({\frac{-r_{ij}}{\lambda_D}}\right) \label{eq:1}
\end{equation}

\noindent where $Q_i$, $Q_j$ are charges on the $i$th and $j$th microparticles, $r_{ij}$ is the separation between them, and $\lambda_D$ is the plasma Debye length. For this reason, dusty plasmas are often treated as a Yukawa system. A molecular dynamics (MD) model of a dusty plasma with Yukawa interactions can capture many features of the dust particles, while still allowing individual particles to be tracked. The same physical processes can therefore be viewed both hydrodynamically (in a fluid description) and kinetically (in a particle description).

Shock waves were traditionally excited in dusty plasmas by a compressional pulse propagating at a supersonic velocity ($v_s> c_l$) relative to the microparticle cloud as a whole. Such shock waves were studied in different plasma configurations using a two-dimensional (2D) monolayer and a three-dimensional (3D) microparticle cloud. The amplitude of these shocks peaked at the beginning of propagation and gradually decreased toward the end. A few experiments have achieved continuously driven shock waves in 3D dusty plasmas, either by flowing the microparticle cloud over a stationary charged wire or by tilting the lower electrode~\cite{10.1063/1.4960032, Jaiswal_2016, PhysRevLett.108.065004}.

Another method to achieve a continuous shock wave in a 2D dust monolayer was demonstrated by Kananovich et al.~\cite{PhysRevE.101.043211}, where a charged exciter wire was moved through a 2D monolayer dusty plasma at a supersonic speed analogous to an exciter piston moving steadily at a supersonic speed in a cylinder of gas. This enabled them to compare and draw the following linear relation between the shock Mach number $M_s=v_s/c_l$ and exciter (piston) Mach number $M_p=v_p/c_l$:

\begin{equation}
    M_s = 1 + sM_p, \label{eq:2}
\end{equation}

\noindent where s is a fit parameter that characterises how the medium compresses. They also measured the shock width ($\delta$) and observed buckling, in which, during shock propagation, some microparticles move above and below the cloud's mean position. In addition, the cloud as a whole can shift vertically downward due to sheath compression. A similar type of shock was studied via 2D molecular dynamics simulation by Lin et al.~\cite{PhysRevE.100.043203}, who modelled a piston with a reflecting boundary condition moving at a constant speed $v_p$ to generate the shock. Thermodynamic and kinetic properties of Shocks in 2D Yukawa systems were studied by M. Marciante and M. S. Murillo~\cite{PhysRevLett.118.025001}.

In most of these simulation studies, periodic or reflecting boundary conditions were used around a 2D simulation box for simplicity, and all microparticles were assumed to be equal in size. This restricted the simulations to capture some of the important features of the particle cloud during shock propagation in an experimental dusty plasma system. The common feature is the buckling of the microparticles under shock compression, which these 2D simulations could not capture. More importantly, since they used a 2D box, their microparticles experienced an infinite vertical potential, which is unrealistic. In the present study, we aim to overcome these discrepancies and bridge the gap between experiments and modelling.

Specifically in this work, we have investigated a piston-driven shock wave in a 2D dusty plasma via MD simulations with realistic parameters, consistent with the experimental result presented by Kananovich et al.~\cite{PhysRevE.101.043211}. We have treated the dusty plasma as a Yukawa One-Component Plasma (YOCP)~\cite{Feng_2016}, set the boundary conditions to match the experiments, and imposed a finite vertical confinement on the microparticle cloud similar to the experimental systems. Our result for the shock propagation speed as a function of exciter speed is in close agreement with the experimental observations by Kananovich et al.~\cite{PhysRevE.101.043211}.


\section{Methodology}  \label{sec: Methods}

Classical Molecular-Dynamics (MD) simulations were performed using LAMMPS~\cite{THOMPSON2022108171} to model piston-driven shock propagation in a finite, strongly coupled 2D dusty plasma. The results presented here used $N=1000$ microparticles in a cubic domain of side length $L=\qty{0.05}{\meter}$ during shock evolution. The simulations were carried out in a 3D Cartesian box, but the microparticles were confined to a quasi-two-dimensional monolayer by a strong vertical restoring force. 

\begin{figure}[!t]
\centering
\begin{tikzpicture}
    \node[anchor=south west, inner sep=0] (img) at (0,0)
    {\includegraphics[width=\columnwidth]{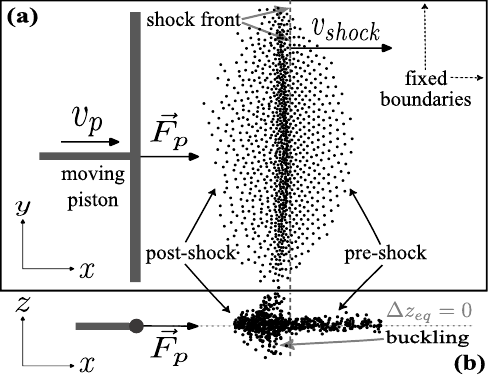}};

    \begin{scope}[x={(img.south east)}, y={(img.north west)}]
        \fill[white] (0.640,0.880) rectangle (0.780,0.965);
        \node[anchor=west, font=\huge] at (0.650,0.925) {$v_s$};
    \end{scope}
\end{tikzpicture}
\caption{Schematic of the piston-driven compressional shock geometry used in the simulations. (a) Top view ($x$-$y$ plane) of the microparticle cloud with a piston (modelled as a coordinate $x_p(t)$ moving with a constant velocity $v_p$ exerting a force $\vec{F}_p$ (see Eq.~\eqref{eq:5}) on the microparticles. The dashed vertical line denotes the shock front generated by the moving piston, and it propagates at a speed $v_{s}$. (b) Side view ($x$-$z$ plane) of the cloud at the same instant, showing the microparticles displacing above and below the $\Delta z_{eq}=0$ plane (shown via the dotted horizontal line).}
\label{fig:1}
\end{figure}

Fixed non-periodic boundaries were imposed along all three directions to avoid artificial particle re-entry through periodic images during shock compression, as shown in Fig. \ref{fig:1}. This accurately replicates the open-system dynamics and finite edge effects of laboratory dusty plasmas.

All simulations were performed in dimensionless Lennard-Jones (LJ) units. The reference mass ($m^*$), length ($\sigma^*$), and time ($\tau^*$) scales were chosen as

\begin{equation}
    m^*=\frac{1}{N}\sum_{i=1}^{N}m_i,\qquad
    \sigma^*=a_{ws},\qquad
    \tau^*=\omega_{pd}^{-1},
    \label{eq:3}
\end{equation}

where $m_i$ is the mass of the $i$th microparticle, $a_{ws}$ is the Wigner-Seitz radius, and $\omega_{pd}$ is the nominal 2D dusty plasma frequency. The corresponding energy scale was $\epsilon^*=m^*\sigma^{*2}/\tau^{*2}$. All the physical parameters were chosen to match the experiment of Kananovich and Goree~\cite{PhysRevE.101.043211}: Coupling parameter $\Gamma=1712$, screening parameter $\kappa=a_{ws}/\lambda_D=1.6$, 2D dusty plasma frequency $\omega_{pd}=\qty{49}{s^{-1}}$, and longitudinal sound speed $c_l=\qty{18.96}{mm.s^{-1}}$. The integration timestep was $\Delta t=\qty{10}{\micro s}$, giving $\Delta t\,\omega_{pd}=4.9\times10^{-4}\ll1$, which is small compared with the characteristic dust collective timescale, i.e., $\Delta t \ll \omega_{pd}^{-1}$. This ensured proper temporal resolution of collective shock-wave dynamics.

\begin{figure}[ht]
    \centering
    \includegraphics[width=\columnwidth]{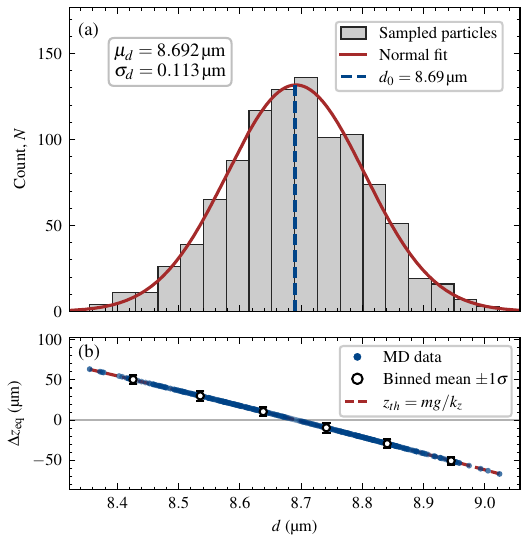}
    \caption{Microparticle size distribution and vertical equilibration data. (a) Histogram of the sampled microparticle diameters ($d$) overlaid with a normal distribution fit with mean diameter $\mu_d = \qty{8.692}{\micro\meter}$ and standard deviation $\sigma_d = \qty{0.113}{\micro\meter}$. The dashed blue line denotes the nominal diameter ($d_0 = \qty{8.69}{\micro\meter}$) set in the simulation. (b) Dependence of time-averaged vertical equilibrium displacement ($\Delta z_{\mathrm{eq}} = z_{\mathrm{eq}}-\langle z_{\mathrm{eq}} \rangle$) on microparticle diameter $d$. Raw simulation data (blue scatter points) show that heavy, large-diameter ($d_i>d_0$) microparticles equilibrate below the mean $z$-level ($\Delta z_{\mathrm{eq}} < 0$). This trend agrees with the mean z-coordinate data for individual bins (open circles) with error bars representing $\pm 1\sigma$ local vertical layer thickness. The red dashed theoretical equilibrium curve derived from vertical force balance ($z_{\mathrm{th}}=-m_ig/k_z$), where $k_z$ is the vertical confinement strength.}
    \label{fig:2}
\end{figure}

The microparticles were modelled as finite-sized spheres with diameters sampled from a normal distribution (as shown in Fig. \ref{fig:2}(a)) with the set diameter $d_0 = \qty{8.69}{\micro\meter}$, consistent with the experimental microparticle size distribution~\cite{PhysRevE.101.043211,10.1063/5.0016504}. All microparticles were assigned the same mass density $\rho=\qty{1510}{\kilo\gram\per\cubic\meter}$ and the same negative charge $Q_d=-Z_de$, with $Z_d\simeq 1.4\times10^4$ and e is the electron charge, as taken from the benchmark experiment~\cite{PhysRevE.101.043211}.

The dust-dust interaction was modelled as the Yukawa (or Debye-Hückel) potential defined in Eq.~\eqref{eq:1} truncated at a cut-off radius $r_c=5\lambda_D$. In LAMMPS, this was implemented using \texttt{pair\_style yukawa}, for which the required screening input is $\lambda_D^{-1}$, rather than the dimensionless screening parameter $\kappa=a_{ws}/\lambda_D$ commonly used in the dusty plasma literature. The Yukawa prefactor $A = Q_d^2/(4\pi\epsilon_0)$ was converted to the corresponding LJ term $A^*=A/(\epsilon^*\sigma^*)$ and then used as the \texttt{pair\_coeff}.

A harmonic confinement potential $U_{\mathrm{conf},i} = \frac{1}{2}(k_x\,x_i^2+k_y\,y_i^2+k_z\,z_i^2)$ was imposed on all microparticles with the corresponding force

\begin{equation}
    \mathbf{F}_{\mathrm{conf},i}
    =-k_x\,x_i\,\hat{\mathbf{x}}
     -k_y\,y_i\,\hat{\mathbf{y}}
     -k_z\,z_i\,\hat{\mathbf{z}}.
    \label{eq:4}
\end{equation}

\noindent This potential $U_{\mathrm{conf},i}$ represented the electrostatic potential well of the plasma sheath, which was balanced by gravity $\mathbf{F}_{\mathrm{g},i}=-m_i g\,\hat{\mathbf{z}}$. This helped levitate the microparticles to a specific z-coordinate, $z_{\mathrm{eq}}=z_{\mathrm{th}}=-m_ig/k_z$. For confining them to a quasi-2D monolayer, the vertical confinement strength was chosen much stronger than the radial confinement strength, i.e., $k_z\gg k_x=k_y$. The microparticles were allowed to equilibrate under these forces following Langevin dynamics. 

The equilibrated cloud was then perturbed by a moving piston represented by a localised repulsive force rather than by a physical wall, as shown in Fig. \ref{fig:1}. The piston coordinate was defined as $x_p(t)=x_p(0)+v_p t$, where $v_p$ is the constant piston speed. The resulting applied piston force applied to the $i$th microparticle ahead of the piston was modelled as a one-sided Gaussian

\begin{equation}
    \mathbf{F}_{\mathrm{p},\,i} = F_0 \exp\!\left[\frac{-(x_i-x_p)^2}{2w^2}\right] \Theta(x_i-x_p)\,\hat{\mathbf{x}}, \label{eq:5}
\end{equation}

\noindent where $F_0$ and $w$ are the amplitude and width of this Gaussian force, and $\Theta$ is the Heaviside function. This force approximates the repulsion produced by a horizontally moving exciter wire while retaining the finite spatial extent of the experimental perturbation. The imposed piston speeds ($v_p$) were supersonic relative to the longitudinal dust-lattice sound speed ($c_l$), and the corresponding piston Mach number was defined as $M_p=v_p/c_l$.

The microparticle dynamics were advanced using the velocity-Verlet algorithm. The equation of motion was

\begin{equation}
    m_i\ddot{\mathbf{r}}_i = -\sum_{i \neq j} \nabla_i \Phi(r_{ij}) + \mathbf{F}_{\mathrm{conf},i}+\mathbf{F}_{\mathrm{g},i} -m_i\nu\dot{\mathbf{r}}_i + \mathbf{F}_{\mathrm{R},i} + \mathbf{F}_{\mathrm{p},i} \label{eq:6}
\end{equation}

\noindent where the first three terms correspond to the conservative forces (Yukawa interaction, confinement force, and gravity), while the fourth term is the neutral gas drag, the fifth term is the random stochastic force (added by the Langevin thermostat in LAMMPS), and the sixth term is the Gaussian piston force. The simulation was done in two phases:

\begin{enumerate}
    \item During equilibration, the piston force was absent, i.e., $\mathbf{F}_{\mathrm{p},i}=0$. The Langevin thermostat applies a damping force with the damping rate $\nu=\nu_L=\qty{2.4}{\hertz}$ and the random force $\mathbf{F}_{\mathrm{R},i}$ satisfying fluctuation-dissipation balance at the target dust temperature $T_d$.
    \item During shock propagation, the piston force was applied as mentioned in Eq. \ref{eq:5}. The Langevin thermostat was switched off, so that $\mathbf{F}_{\mathrm{R},i}=0$, and the dissipative term represented Epstein neutral-gas drag with $\nu=\nu_{dn}=\qty{2.4}{\hertz}$. This $\nu_{dn}$ represented the dust-neutral collision frequency calculated for the gas pressure taken from the experimental work by Kananovich et al.~\cite{PhysRevE.101.043211}.
\end{enumerate}

\begin{figure}[ht]
    \centering
    \includegraphics[width=\columnwidth]{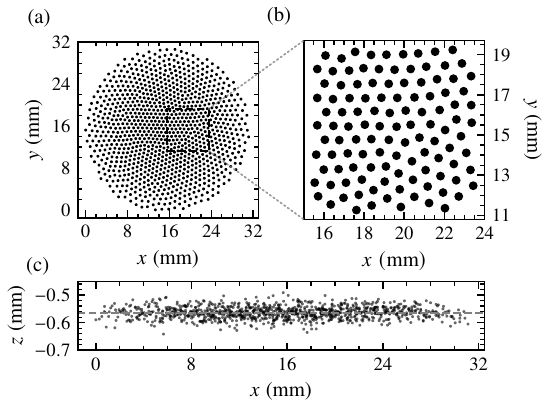}
    \caption{Pre-shock 2D equilibrated microparticle cloud with $\Gamma=1712$ and $\kappa=1.6$ at a time $t=t_0$. (a) Top view ($x$-$y$ plane) of the entire equilibrated cloud comprising $N=1000$ microparticles, where the dashed rectangle delineates the magnified ROI. (b) Magnified Top view showing the strongly-coupled hexagonal lattice with lattice constant $b=\qty{0.825}{\milli\meter}$ calculated using $\mathrm{RDF}$ before shock excitation. (c) Side view ($x$-$z$ plane) of the entire cloud showing the microparticles levitating under the balance of the gravitational force ($\mathbf{F}_{\mathrm{g},i}=-m_ig\ \hat{z}$) and the vertical sheath electric field ($\mathbf{F}=-k_zz_i\ \hat{z}$). The horizontal dashed line denotes the theoretical mean equilibrium levitation height ($z_{\mathrm{eq}}=\qty{-0.563}{\milli\meter}=-mg/k_z$). The narrow vertical dispersion (range of z-coordinates is $\Delta z_{\mathrm{eq}}=\qty{0.15}{\milli\meter}$) here confirms the formation of a 2D monolayer showing strong vertical confinement such that $\Delta z_{\mathrm{eq}}<b$.}
    \label{fig:3}
\end{figure}

The obtained structure was analysed from saved microparticle coordinates. For each saved frame, the cloud was divided into vertical bins of width $\Delta x = \qty{1}{\milli\meter}$. The areal number density in each bin was computed as

\begin{equation}
    n(x,t)=\frac{N_{\rm bin}(x,t)}{\Delta x\,L_y}, \label{eq:7}
\end{equation}

\noindent where \(N_{\rm bin}(x,t)\) is the number of microparticles in the bin and $L_y = \qty{10}{\milli\meter}$ is the transverse extent of the analysed region.

The shock-front position was defined as the position $x_s(t)$ at which the areal number density profile $n(x,t)$ reached its maximum within the analysis window. Frames in which the density profile contained multiple comparable peaks or boundary-affected structures were excluded from the fit. A linear fit:

\begin{equation}
    x_s(t)=x_s(0)+v_st,
    \label{eq:8}
\end{equation}

\noindent yielded the shock speed $v_s$, and the shock Mach number was calculated as $M_s=v_s/c_l$. Out-of-plane motion was quantified using $\Delta z_i=z_i-z_{eq}$ to identify buckling during shock passage.

\section{Results and Discussion}

\begin{table}[ht]
\renewcommand{\arraystretch}{1.3}
\caption{Comparison of Simulated and Experimental Structural Parameters for the Unperturbed Microparticle Cloud ($\Gamma = 1712$, $\kappa = 1.6$). \label{tab:1}}
\centering
\begin{tabular}{|c|c|c|}
\hline
\textbf{Parameter} & \textbf{Simulation data} & \textbf{Experimental data~\cite{PhysRevE.101.043211}} \\
\hline
$b$ & $\qty{0.825}{\milli\meter}$ & $\qty{0.80}{\milli\meter}$ \\ 
$n_0$ & $\qty{1.69}{\milli\meter^{-2}}$ & $\qty{1.81}{\milli\meter^{-2}}$ \\
$v_{\mathrm{rms}}$ & $\qty{0.598}{\milli\meter\per\second}$ & $\qty{0.499}{\milli\meter\per\second}$\\
\hline
\end{tabular}
\end{table}

\begin{figure*}[ht]
    \centering
    \includegraphics[width=0.85\textwidth]{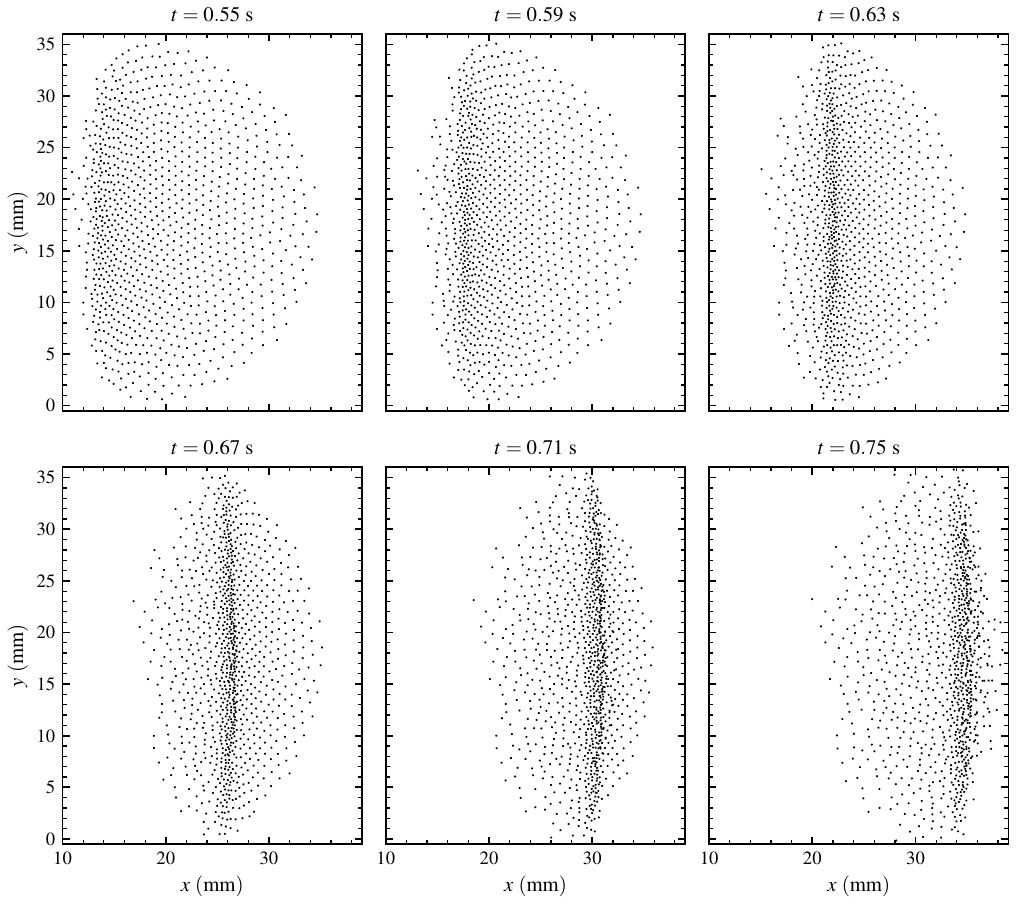}
    \caption{Time evolution and propagation of the compressional pulse for $v_p = \qty{101.6}{\milli\meter\per\second}$ as seen in the top-view ($x$-$y$ plane) of the entire microparticle cloud. The pulse propagates in the $+x$ direction, excited by the force $\mathbf{F}_p$ exerted by the moving piston. The compressional region develops a sharper leading edge as it propagates.}
    \label{fig:4}
\end{figure*}

After the microparticles formed an equilibrated monolayer (as shown in Fig. \ref{fig:3}), some structural parameters of this monolayer were calculated from the simulation data, and are shown in Table \ref{tab:1}. The unperturbed areal number density ($n_0$) of the equilibrated bulk of the microparticle cloud was obtained using the binning process described in Sec. \ref{sec: Methods}. The lattice constant ($b$) was calculated using the radial distribution function ($\mathrm{RDF}$), and the root mean square speed $v_{\mathrm{rms}}$ was calculated for the last 5 seconds of the equilibration phase.

Because the microparticles had a finite mass distribution ($m_i$), their equilibrium levitation heights ($z_{\mathrm{eq}}=-m_ig/k_z$) were not identical. While a 1.26\% coefficient of variation ($\sigma_d/\bar{d}$) in diameter is small, this size variance was explicitly modelled because perfectly identical microparticles in an MD simulation artificially force a defect-free hexagonal crystal. Laboratory dusty plasma experiments generally use industrially manufactured microparticles that have slight polydispersity due to limitations of the manufacturing process. This induces small-scale topological defects and variations in the equilibrium levitation height ($z_{\mathrm{eq}}$) of individual microparticles, as shown in Fig. \ref{fig:2}(b). However, this variation in the levitation heights ($\Delta z_{\mathrm{eq}}$) is much smaller than their average inter-particle distance ($b$), meaning that the microparticles were in a quasi-2D monolayer.

\begin{figure}[ht]
    \centering
    \includegraphics[width=0.4\textwidth]{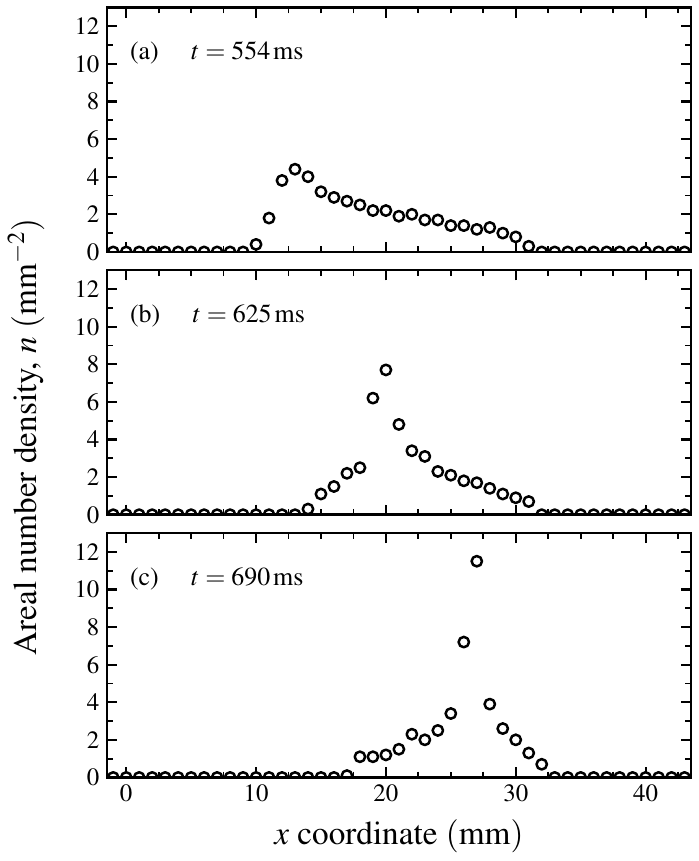}
    \caption{Areal number-density profiles at three times for $v_p = \qty{101.6}{\milli\meter\per\second}$}
    \label{fig:5} 
\end{figure}

Since the experimental runs excited multiple waves at different piston speeds ($v_p$) ranging from $\qty{63.5}{\milli\meter\per\second}$ to $\qty{101.6}{\milli\meter\per\second}$, multiple simulation runs were conducted at these piston speeds. The compressional pulse obtained during the cloud perturbation via the force exerted by the piston (see Eq.~\eqref{eq:5}) is shown in Fig. \ref{fig:4}. The word \textit{compressional} highlights the observation that the microparticle number densities increased manifold in the localised region of this pulse, as seen in the time evolution of this number density plotted in Fig. \ref{fig:5}. This pulse was identified as a shock wave because of the following observations:

\begin{itemize}
    \item Within a short time interval of 0.1 seconds, the particle areal number density value peaked. And this peak's leading edge was much steeper than its trailing edge. This is visible in both Fig. \ref{fig:4} and Fig. \ref{fig:5}.
    \item The obtained pulse speeds $v_s$ were greater than $\qty{70}{\milli\meter\per\second}$ in all four runs, compared to the sound speed $c_l = \qty{18.96}{\milli\meter\per\second}$. Thus, the pulse's propagation was highly supersonic ($v_s>c_l$). For our simulations, the maximum Mach number for the simulation pulse was $M_s = 5.8$, while that obtained in the experiment~\cite{PhysRevE.101.043211} was $M_s=6.2$.
    \item The maximum compression ratio $n_{\mathrm{peak}}/n_0$ obtained for our pulse ranged from 4.5 to 7.5 for all the runs analysed. Such high number densities implied that the pulse's amplitude was large and therefore highly nonlinear.
\end{itemize}

\begin{figure}[ht]
    \centering
    \includegraphics[width=\columnwidth]{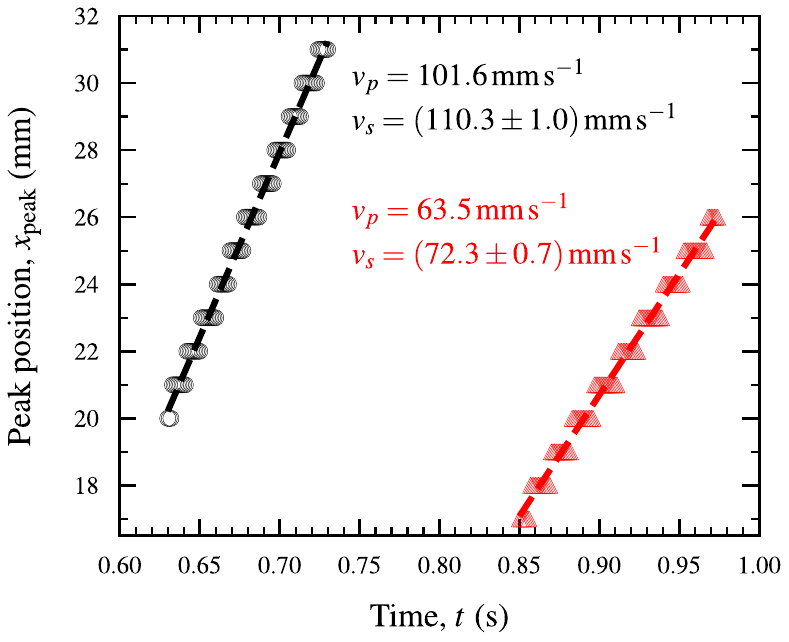}
    \caption{Representative plot of the time series of the position of the pulse’s peak, for two runs. The data points are the positions ($x_{\mathrm{peak}}$) of the peak areal number density, obtained from simulation data. The dashed black line represents a linear fit, yielding a shock speed $v_s = \qty{110.3}{\milli\meter\per\second}$ for $v_p=\qty{101.6}{\milli\meter\per\second}$, while the dashed red line represents the same for $v_p=\qty{63.5}{\milli\meter\per\second}$ giving a shock speed $v_s = \qty{72.3}{\milli\meter\per\second}$. The quoted uncertainties are one-standard-error uncertainties obtained from the linear least-squares fits.}
    \label{fig:6}
\end{figure}

The pulse's propagation speed $v_s$ was obtained as the slope in a plot of the peak areal number density positions ($x_{\mathrm{peak}}$) as a function of time. The representative examples of these plots are shown in Fig. \ref{fig:6}. For the two piston speeds for which exact experimental values are explicitly available from the peak-tracking plot (Fig. 4 in~\cite{PhysRevE.101.043211}), the present simulations give $v_s = (72.28 \pm 0.73)\,\si{\milli\meter\per\second}$ for $v_p = \qty{63.5}{\milli\meter\per\second}$ and $v_s = (110.26 \pm 0.99)\,\si{\milli\meter\per\second}$ for $v_p = \qty{101.6}{\milli\meter\per\second}$, compared with the experimental values of $(83.7 \pm 0.8)\,\si{\milli\meter\per\second}$ and $(117.5 \pm 1.1)\,\si{\milli\meter\per\second}$ respectively.

\begin{figure}[ht]
    \centering
    \includegraphics[width=\columnwidth]{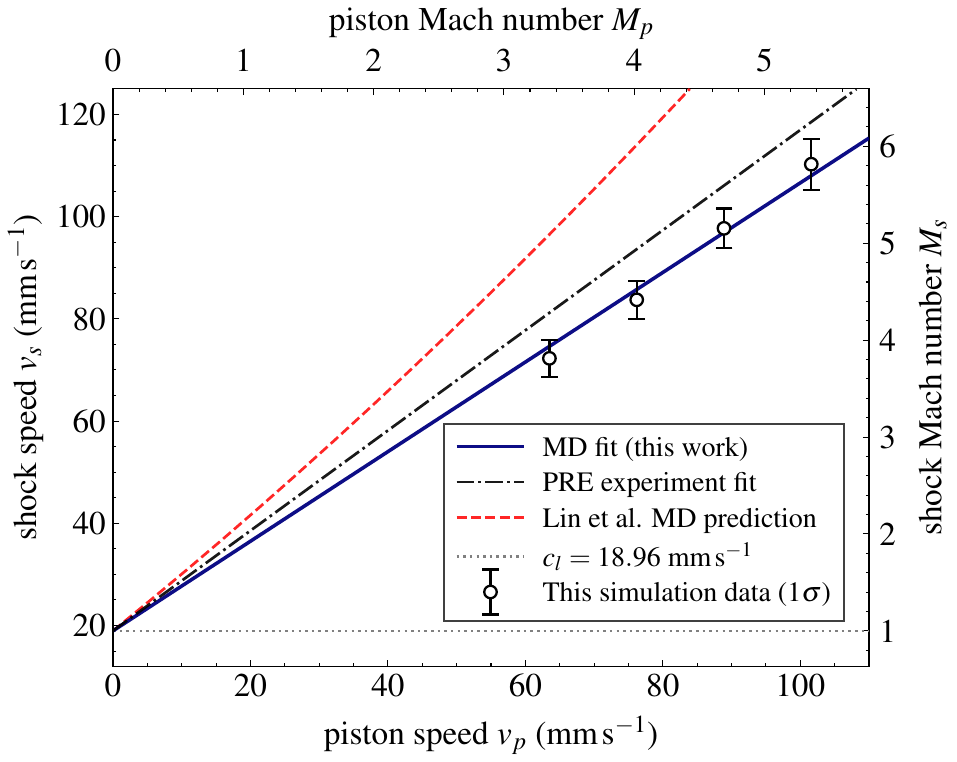}
    \caption{A comparison of shock speed ($v_s$) results obtained in this work, in experiments~\cite{PhysRevE.101.043211}, and in earlier MD simulations done by Lin et al.~\cite{PhysRevE.100.043203}. The present MD data are fitted using Eq.~\eqref{eq:2}, giving $s=0.88$, while the PRE experiment~\cite{PhysRevE.101.043211} obtained $s=0.98$.}
    \label{fig:7}
\end{figure}

The main result of the experiments by Kananovich et al.~\cite{PhysRevE.101.043211} was the generation of a driven shock wave via a continuously moving exciter in a 2D dusty plasma and obtaining the dependence of the shock speed $v_s$ on the exciter (piston) speed $v_p$. The shock speed's monotonic increase with the piston speed was empirically fitted to Eq.~\eqref{eq:2} with a slope of $s=0.98$. A similar analysis was performed on the simulation data obtained in this work, yielding a slope of $s=0.88$. This comparison is shown in Fig. \ref{fig:7}, where both the experiment data as well as results from a previous MD simulation done by Lin et al.~\cite{PhysRevE.100.043203} are compared.

The simulation data reproduce the monotonic increase of $M_s$ with $M_p$ and are approximately described by the empirical relation in Eq.~\eqref{eq:2}. The observed compressional pulses exhibit many features of a shock wave and are supersonic over the entire range. Besides, the simulation data points fit the linear curve shown in Fig. \ref{fig:7} well, providing confidence in the calculated shock speeds and the empirical relation in Eq.~\eqref{eq:2}. Within the present range of piston speeds and uncertainties, the additional quadratic correction proposed by Lin et al.~\cite{PhysRevE.100.043203} is not required to describe the present data.

However, the simulated shock speeds are lower than the experimental values, especially at lower piston speeds. This discrepancy may arise because the piston is modelled as a moving Gaussian force field rather than a physical charged wire with an actual plasma sheath. Other possible sources include assumptions made to simplify the model, such as a constant microparticle charge, a constant parabolic confinement potential, and the absence of ion-wake effects.

A central feature of the simulated shock waves is the out-of-plane response of the microparticles in the compressed region, a phenomenon known as \textbf{buckling}~\cite{sheridan2010power,zampetaki2020buckling}. As the piston compresses the monolayer, the strong in-plane Yukawa repulsion drives a fraction of the microparticles above and below the equilibrium plane $\Delta z_{eq}=0$, rather than being further compressed within it. This is clearly visible in the side view of Fig.~\ref{fig:1}(b), in which the pre-shock cloud remains tightly confined to the monolayer while the post-shock region exhibits a pronounced vertical scatter of particles about $\Delta z_{eq}=0$. Such out-of-plane motion under shock compression was previously reported in the 2D dusty plasma experiment of Kananovich and Goree~\cite{10.1063/5.0016504}; however, to the best of our knowledge, it has not been reproduced in any prior MD simulation of 2D dusty plasma shocks. Earlier 2D MD studies~\cite{PhysRevE.100.043203,PhysRevLett.118.025001} confined the microparticles strictly to a plane, effectively imposing an infinite vertical potential, so the out-of-plane response of the lattice to strong compression could not emerge. The present simulations recover it because the microparticles are evolved in a 3D Cartesian domain with a finite, harmonic vertical confinement that mimics the experimental conditions. Reproducing buckling in a piston-driven 2D dusty plasma shock thus closes a long-standing gap between experimental observations and MD modelling, and is one of the principal results of this work.

A separate, more incidental feature was the appearance of a small number of microparticles that travelled significantly ahead of the shock front. These might be the pre-heated particles previously reported in past MD simulations of shocks in 2D dusty plasmas~\cite{PhysRevLett.118.025001}.

\section{Conclusion and Future Outlook}

In conclusion, we performed MD simulations of continuously driven shock waves in a two-dimensional dusty plasma. Compared with earlier modelling efforts, our approach incorporates a three-dimensional simulation domain with finite vertical confinement, a realistic distribution of microparticle sizes, and fixed (non-periodic) boundaries that mimic the experimental cloud confinement. These features enable close agreement with previously reported experimental results, including out-of-plane buckling of the shock-loaded crystal. This phenomenon was observed experimentally but, to the best of our knowledge, not reproduced in prior MD simulations of shocks in 2D dusty plasmas.

This framework can be extended to investigate additional out-of-plane and wake-mediated physics by incorporating ion flow and the resulting non-reciprocal inter-particle interactions.

\section*{Acknowledgments}
Appalachian State University was supported by the United States Department of Energy under Grant No. DE-SC0025444 and the National Science Foundation under Grant No. PHY-2510502. A.K. would like to thank Prof. J.~Goree for insightful discussions. S. J. acknowledges the support from the IISER Pune startup fund.
\begin{IEEEbiographynophoto}{Prateek Lamoria}
Prateek Lamoria is an undergraduate student in Indian Institute of Science Education and Research (IISER) Pune India.
\end{IEEEbiographynophoto}
\begin{IEEEbiographynophoto}{Anton Kananovich}
 is currently an assistant professor in Department of Physics and Astronomy at Appalachian State University Boone, NC. His research focuses on shock wave dynamics in dusty plasma.
\end{IEEEbiographynophoto}
\begin{IEEEbiographynophoto}{Surabhi Jaiswal}
is currently an assistant professor in physics department at Indian Institute of Science Education and Research (IISER) Pune India. Her research focuses on flow induced nonlinear waves and instability in complex plasma.
\end{IEEEbiographynophoto}
\bibliographystyle{IEEEtran}

\end{document}